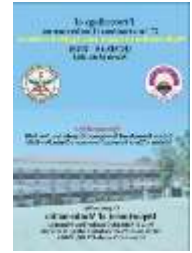

*RESEARCH ARTICLE*

# Cryptanalysis on Secure ECC based Mutual Authentication Protocol for Cloud-Assisted TMIS

**Diksha[1*], Meenakshi[2]**

[1*,2]Srinivasa Ramanujan Department Of Mathematics, Central University Of Himachal Pradesh, Dharamshala, 176215, Himachal Pradesh, India.

*Corresponding Author E-mail: **dikshachouhan1035@gmail.com**

**ABSTRACT:**
The creation of TMIS (Telecare Medical Information System) makes it simpler for patients to receive healthcare services and opens up options for seeking medical attention and storing medical records with access control. With Wireless Medical Sensor Network and cloud-based architecture, TMIS gives the chance to patients to collect their physical health information from medical sensors and also upload this information to the cloud through their mobile devices. The communication is held through internet connectivity, therefore security and privacy are the main motive aspects of a secure cloud-assisted TMIS. However, because very sensitive data is transmitted between patients and doctors through the cloud server, thus security protection is important for this system. Recently, Kumar et al designed a mutual authentication protocol for cloud-assisted TMIS based on ECC [2]. In this paper, we revisited this scheme and traced out that their scheme has some significant pitfalls like health report revelation attack, and report confidentiality. In this study, we will provide the cryptanalysis of the scheme developed by Kumar et al.

## 1. INTRODUCTION:

With the advancement in technology, patients can receive care from medical professionals through the internet. The inaccessibility of remote locations and unavailability of facilities makes modern healthcare facilities difficult and these healthcare facilities are as expert advice, proper diagnosis, clinical tests, etc. Due to poor return on investments, no one is interested to invest in these areas and also doctors are not interested to serve in those areas which are under development. This leads the patients to have to travel long distances and spend a lot of money to get medical treatment. Some patients leave their hopes on their fates or lived with treatments from local health workers. In this case, a platform Telecare Medical Information System (TMIS) facilitates the patients and doctors with the communication between them and provides medical assistance in the patient's home. At the moment, everyone is paying close attention to cloud computing. It exhibits a significant potential for providing medical services online in TMIS due to its profitable specialties including on-demand self-service, more resilience, and resource sharing.

TMIS can gain various financial and functional advantages from cloud-based architecture, including flexible medical data storage, lower costs, better accessibility, and higher standards of care. But it also faces a lot of problems like reliability, privacy, security, and many others. Since the patient's data is transmitted between the entities over an insecure public channel. Therefore data security, confidentiality, and also its authenticity are major priorities in cloud-based TMIS.

The likelihood of a mischievous insider for consumers of cloud services is increased by the consolidation of IT services and clients under one management domain and the general lack of transparency into provider methods and procedures. A provider could choose to keep secrets about how it supervises personnel and provides them access to physical and digital resources, analyses, and reports on quality management. To further complicate matters, hiring standards and procedures for cloud staff are frequently hidden or not disclosed at all. This type of circumstance presents an interesting opportunity for an enemy, who may be engaged in a nation-state-sponsored attack, organized crime, hobbyist hackers, or even industrial crime. Such an adversary might be able to gather sensitive information or take total control of the cloud services thanks to the level of access allowed with little to no chance of being detected [3].

A safe cloud-assisted TMIS has required some important characteristics such as:
a) **Message authentication:** The system's users must be able to confirm that a message delivered over an unsecured channel was actually sent to and received by a legitimate recipient without interference from an unauthorised third party.
b) **Patient anonymity:** Any of the messages on the public channel should not express the true identity of the patient and anyone can't estimate and find the patient's true identity.
c) **Patient unlinkability:** Any outsider can't estimate the relationship between patient and doctor.
d) **Report confidentiality:** The sensitive data of the patient should be accessible by only the appointed doctor.
e) **Non-repudiation:** The patient, hospital, and doctor cannot contest the veracity of the digital signature they placed on a document or the message they sent.

## 2. RELATED WORKS:
In 2012, Patra et al proposed a cloud-based model for making a rural healthcare information system [4]. Furthermore, Chen et al. put forth a cloud-based plan for exchanging medical data. They use the pairing technology with an asymmetric key to encrypt the information in this scheme [5]. But later on, this scheme faced some problems [6]. In 2015, an improved patient-server mutual authentication protocol for TMIS was proposed by Amin et al. [7]. This protocol concerns a biometric-based remote user authentication scheme for TMIS. Wu et al also propose a mutual authentication scheme for this healthcare application [8]. In 2014, Wen et al introduced an anonymous authentication scheme for TMIS [9]. Further, Xu et al provided a secure and effective two-factor mutual authentication and key agreement mechanism for TMIS based on the ECC [10]. Moreover, with the help of WMSN, He et al built a strong anonymous authentication technique for healthcare applications [11]. Later on, a symmetric key-based authentication method for wireless medical sensor networks was proposed by Jangirala et al [12]. Chen et al also put forward a cloud-assisted secure authentication technique for healthcare systems [13]. To address these issues with Chen et al's method, Chiou et al offered a modified authentication scheme in 2016. Chiou et al, however, noted that this framework does not provide message authentication, patient anonymity, etc [14]. Further, Mohit et al reviewed [14] and offered a mutual authentication framework based on cloud for healthcare systems [15] due to some pitfalls in the scheme [14]. Also, a cloud-assisted effective mutual authentication protocol for healthcare systems was introduced by Kumar et al [16]. For the TMIS environment, Li et al proposed a cloud-assisted mutual authentication framework with preserved privacy [1]. In 2018, Kumar et al reviewed Li et al's scheme and presented a framework for cloud-assisted TMIS with mutual authentication using ECC [2]. But now, we reviewed this protocol and established some design flaws like health report revelation attack, and report confidentiality.

The remaining paper is split up into different sections as follows:
Section 3 represents the review of the scheme [2] and the difficulties which are faced by this scheme described in section 4. Lastly, section 5 discusses the conclusion of this paper.

## 3. REVIEW OF KUMAR *ET AL*. SCHEME[2]
A mutual authentication and preserved privacy framework for cloud-assisted TMIS is offered by Kumar et al. In their scheme, total five important roles of bodies take place. This programme is divided into four stages: the upload phase for healthcare centre, the patient data upload phase, the treatment phase, and the checkup phase.
1. H uploads the inspection medical report of P in C in the Healthcare Centre Upload Phase (HUP).
2. In the Patient data upload phase (PUP), P uploads his/her current medical report from the embedded body sensor to C.
3. In the treatment phase (TP), D will recommend treatments to C for P of the appropriate body.
4. P obtains the report from C during the Checkup Phase (CP), as directed by D.

**Notations:**

| $ID_x$ | Entity x's unique identity |
|---|---|
| NID | Dynamic pseudo random number |
| $sn_x$ | Serial number of $x^{th}$ participant |
| $PR_x$ | Private key of x |
| $PK_x$ | Public key of x |
| A | Adversary |
| $m_H$ | Inspection report of P generated by H |
| $S_i(M)$ | Using key i to sign M |
| $m_B$ | Health report of P from body sensor |
| h(.) | Hash function |
| $m_D$ | Medical report of P generated by D |
| $SK_{xy}$ | Session-key between x and y |
| $K_x$ | Computing key of x entity |
| G | Elliptic curve group (additive) |
| G | Base point of G |
| $Sig_x$ | Signature of entity x |
| $Z_p^*$ | Additive group of large prime of order p |
| $V_i(M)$ | Using key i to verify M |

**HUP:**

The registration of P takes place in H, where an NID is allotted or assigned to P by H with the help of mobile device securely. P's inspection report $m_H=(ID_P, data_P)$ is uploaded (in C) by H after mutual authentication between H and C in this phase and this happen with following procedure:

1.) Initially, H sends a massage which consist of $ID_H$, random number a (from $Z_p^*$) and $T_{H1}$ to C through secure channel.

2.) After getting this message, C checks the validity of $T_{H1}$ with $T_{C1} - T_{H1} \leq \Delta T$. Then a random number b ($\epsilon\ Z_p^*$) is generated by C and also C computes $S_1= h(ID_H||a||b||T_{H1})$, $K_1= h(ID_H||a||T_{H1})$ which is used for the encryption of the (b, $S_1, T_{C2}$) to get $E_1$. After that, C sends ($E_1, T_{C2}$) to H via public channel.

3.) When this message is collected by H, then firstly H verifies $T_{H2}-T_{C2} \leq \Delta T$ (if not, then the session will be terminated) and ready the key $K_1'$ to decrypt the $E_1$. Where $K_1'= h(ID_H||a||T_{H1})$ and $E_1= E_{K1}(b, S_1, T_{C2})$. After that, H computes $S_1'=h(ID_H||a||b||T_{H1})$ and verifies that $S_1'=?\ S_1$. Then H computes $SK_{HC} = h(ID_H||S_1'||abg||T_{C1})$ and key $K_2 =h(ID_P||ID_H||NID)$ to encrypt $m_H$ i.e. $C_H = E_{K2}(m_H)$. Next, H makes digital signature $Sig_H = S_{PRH}(h(m_H))$, $S_2 = h(SK_{HC}||C_H||Sig_H||T_{H3})$ and encrypts $E_2 = E_{SKHC}(ID_P, S_2, C_H, NID, Sig_H, T_{H3})$. Sends the message ($E_2, T_{H3}$) to C via public channel.

4.) Then C verifies $T_{C3}-T_{H3} \leq \Delta T$ after collecting the message from H and computes the session key $SK_{CH} = h(ID_H||S_1||abg||T_{C1})$ to decrypt $E_2$. Next, C computes $S_2' = h(SK_{CH}||C_H||Sig_H||T_{H3})$ and check whether $S_2'= S_2$ or not. If not then session will stop there otherwise C stores NID, $ID_P$, $Sig_H$, $C_H$.

**PUP:**

The embedded sensor in the patient's body collects the health information $m_B= (ID_P, data_B)$ and securely sends this information to the patient's mobile phone. C gives the sequence number $sn_x$ and $m_H$ to P after making the request by P (using his/her $ID_P$ and NID) to C.

1.) P receives a health report from the embedded body sensor in form of $m_B$ and sends ($ID_P$, NID, $T_{P1}$) to C through the trustable channel.

2.) After receiving this message, C verifies $T_{C4}-T_{P1} \leq \Delta T$. Next, C computes $I = sn_x + h(NID||ID_P)$ and generates the random number c ($\epsilon\ Z_p^*$). Then C computes the hash value $S_3 = h(NID||ID_P||C_H||Sig_H||c||T_{C5})$, encrypts the message ($Sig_H$, $C_H$, $S_3$, $ID_H$, c, $T_{C5}$) using the $sn_x$ and gets $E_3$. Further, C sends ($E_3$, I, $T_{C5}$) to P via a public channel.

3.) P verifies $T_{P2}-T_{C5} \leq \Delta T$ after getting the message ($E_3$, I, $T_{C5}$) from C and computes $Y = I + h(NID||ID_P)$ to decrypt $E_3$. After that, P computes $S_3'= h(NID||Y||C_H||Sig_H||c||T_{C5})$ and verifies whether $S_3' = S_3$ or not. The session will end there if it does not. Otherwise, P generates random number d ($\epsilon\ Z_p^*$) and computes the session key $SK_{PC} = h(ID_P||ID_H||C_H||S_3'||cdg||T_{C5})$. Now, P computes a key $K_3 = h(ID_P||ID_H||NID)$ for decryption of $m_H^* = D_{K3}(C_H)$ and checks $m_H^* = m_H$. Furthermore, P verifies whether $V_{PKH}(Sig_H) = h(m_H)$ or not. After verification, P computes $K_4$ $= h(ID_P\|ID_D\|Y)$ and encrypts $C_P = E_{K4}(m_H, m_B)$. Moreover, P makes the digital signature $Sig_P = S_{PRP}(h(m_B))$, computes the hash value $S_4 = h(SK_{PC}\|C_P\|Sig_P\|S_3'\|cdg\|T_{P3})$ and using Y as a key to encrypt $E_4 = E_Y(d, S_4, Sig_P, C_P, T_{P3})$. Lastly, through a public channel, P communicates the message $(E_4, T_{P3})$ to C.

4.) On accepting this message, C checks $T_{C6}-T_{P3} \leq \Delta T$. If hold, then C decrypts $E_4$ with $sn_x$, computes $SK_{CP} = h(ID_P\|ID_H\|C_H\|S_3\|cdg\|T_{C5})$ and computes $S_4' = h(SK_{CP}\|C_P\|Sig_P\|S_3\|cdg\|T_{P3})$ to check whether $S_4 = S_4'$ or not. In that case, C ends the session. If not, C verifies P and saves $C_P$, $ID_P$, and $Sig_P$ in the database.

**TP:**
The authentication establishes between D and C in this phase. D takes the report of the patient from C for diagnosis. After that D uploads the treatment report $m_D = (ID_P, data_D)$ for the respective patient in C.

1.) D sends a message $(ID_D, r, T_{D1})$ to C through a secure channel. (Where r is the generated random number.)
2.) After collecting this message, C firstly verifies $T_{C7}-T_{D1} \leq \Delta T$, computes $J=sn_x + h(ID_D\|r)$, generates the random number s, computes the hash value $S_5 = h(ID_P\|ID_D\|Sig_H\|Sig_P\|C_P\|T_{C8})$, encrypts $(Sig_P, Sig_H, ID_P, NID, C_P, s, S_5, T_{C8})$ using $sn_x$ (i.e. $E_5$) and sends the message $(E_5, J, T_{C8})$ to D through the insecure channel.
3.) D verifies $T_{D2}-T_{C8} \leq \Delta T$ after receiving this message and computes $Z = J + h(ID_D\|r)$ to decrypt $E_5$ and gets $(NID, ID_P, Sig_P, Sig_H, s, S_5, C_P, T_{C8})$. After that, D computes $S_5' = h(ID_P\|ID_D\|Sig_H\|Sig_P\|C_P\|T_{C8})$ and checks $S_5' =? S_5$. After this verification D successfully authenticates C and computes the key $K_5 = h(ID_P\|ID_H\|NID)$. Moreover, D uses this key $K_5$ to decrypt $C_P$ and gets the P's reports $m_H$, $m_B$. Further, D performs $V_{PKH}(Sig_H) = h(m_H)$ and $V_{PKP}(Sig_P) = h(m_B)$. If it holds, then D diagnoses these reports and makes treatment report $m_D$, using key $K_5$ to encrypt $C_D = E_{K5}(m_H, m_B, m_D)$. Next, D makes his/her digital signature $Sig_D = S_{PRD}(h(m_D))$ and computes $S_6 = h(ID_P\|ID_D\|C_D\|Sig_D\|Sig_P\|T_{D3})$. Also, D computes session key $SK_{DC} = h(S_6\|ID_P\|ID_D\|Sig_D\|Sig_P\|rsg\|T_{D3})$, encrypts $E_6 = E_Z(Sig_D, C_D, S_6, T_{D3})$ and sends the message $(E_6, T_{D3})$ to C through a public channel.
4.) Upon receiving this message, C verifies $T_{C9}-T_{D3} \leq \Delta T$ and decrypts $D_{snx}(E_6)$. Furthermore, C computes $S_6' = h(ID_P\|ID_D\|C_D\|Sig_D\|Sig_P\|T_{D3})$ and verifies $S_6' =? S_6$. C authenticates D after only this successful verification, now C computes $SK_{CD} = h(S_6'\|ID_P\|ID_D\|Sig_D\|Sig_P\|rsg\|T_{D3})$. Lastly C stores $C_D$ and $Sig_D$ in its database.

**CP:**
After Treatment Phase, P collects his/her encrypted report $m_D = (ID_P, data_D)$ from C after mutual authentication between them.

1.) Firstly, P sends the message $(ID_P, NID, x, sn_x, T_{P4})$ to C through the secure channel. (Where x is the random number taken from $Z_p^*$).
2.) On accepting this message, C verifies $T_{C10}-T_{P4} \leq \Delta T$ and generates a random number y. Next, C computes $S_7 = h(SK_{CP}\|ID_P\|ID_D\|C_D\|xyg\|Sig_P\|T_{C11})$ and with the help of session key, C encrypts $E_7 = E_{SKCP}(ID_D, Sig_D, C_D, S_7, y, T_{C11})$. Lately, C sends $(E_7, T_{C11})$ to P through the public channel.
3.) P verifies $T_{P4}-T_{C11} \leq \Delta T$ and proceeds with the decryption of $E_7$ using the session key. Now P compute $S_7' = h(SK_{PC}\|ID_P\|ID_D\|C_D\|xyg\|Sig_P\|T_{C11})$ and verifies $S_7' =? S_7$. P authenticates C after only this successful verification and decrypts the report $C_D$ using $K_4 = h(ID_P\|ID_D\|Y)$. Now P collects all his/her reports $m_H$, $m_B$, $m_D$ and verifies the digital signature with the help of the public key of D as $V_{PKD}(Sig_D) =? h(m_D)$. If verification holds, then P again encrypts reports(say $C_E$) with the help of the same key $K_4$ such as $C_E = E_{K4}(m_H, m_B, m_D)$, and computes $S_8=h(SK_{PC}\|S_7'\|C_E\|Sig_P\|Sig_D\|xyg\|T_{P6})$. Moreover, P encrypts $E_8 = E_{SKPC}(C_E, S_8, T_{P6})$ and uses the public channel to deliver the message $(E_8, T_{P6})$ to C.
4.) After collecting the message, C verifies $T_{C12}-T_{P5} \leq \Delta T$ and decrypts $E_8$ using session key $SK_{CP}$. Further, C computes $S_8' = h(SK_{CP}\|S_7'\|C_E\|Sig_P\|Sig_D\|xyg\|T_{P6})$ and checks $S_8 =? S_8'$. C authenticates P after only this successful verification and stores $C_E$ in its database.

## 4. THE CRYPTANALYSIS OF KUMAR
**Et al's scheme:**
### 4.1 Health report revelation attack:
#### 4.1.1. Medical report revelation attack in HUP:
When HUP starts, H sends its identity, $ID_H$, and random number m from $Z_p^*$ and sends $ID_H$, a, and $T_{H1}$ to cloud server C through the trustable channel. In this case, H is simply sending its own identity.
And after the third step, H sends $E_2$ together with $T_{H3}$ to C via insecure channel, where $E_2$ is encrypted by session key $SK_{HC}$ as follows $E_2 = E_{SKHC}(ID_P, NID, S_2, C_H, Sig_H, T_{H3})$, $SK_{HC}=h(ID_H\|A_1'\|abg\|T_{C1})$, $K_2=h(ID_P\|ID_H\|NID)$, $C_H = E_{K2}(m_H)$.

When C receives all these things, then C checks the time stamp and computes $SK_{HC}=h(ID_H\|S_1'\|abg\|T_{C_1})$, decrypts $E_2$ with this session key. From this action of C, C gets $ID_P$, NID, $C_H$, $S_2$, $Sig_H$, and $T_{H3}$.

Now the privileged insider of C, (let's say A) can compute $K_2$, which was used to encrypt $m_H$, and also used for the decryption of $C_H$ because A has $ID_P$, NID, $ID_H$. Thus A can decrypt $m_H = D_{K2}(C_H)$, which contains the inspection report of the patient and the $ID_P$.

### 4.1.2. Medical report revelation attack in PUP:

After the third step in PUP, P sends $E_4$ together with $T_{P3}$ to C via an insecure channel, where $E_4$ is encrypted by sequence number $Y (= sn_x)$ as follows $E_4 = E_Y(d, S_4, Sig_P, C_P, T_{P3})$, and $C_P = E_{K4}(m_H, m_B)$ is encrypted by P with the help of key $K_4 = h(ID_P\|ID_D\|Y)$. Whereas in TP, D decrypts $(m_H, m_B) = D_{K5}(C_P)$ with $K_5 = h(ID_P\|ID_H\|NID)$, and this key $K_5$ is the same as $K_2$.

But as in 4.1.1, A already got the key $K_2$, and using this key he/she can decrypt $C_P$. Hence, the inspection report $m_H$ together with the health report $m_B$ (generated by sensors) will be revealed by A in PUP.

### 4.1.3. Medical report revelation attack in TP:

Similarly, after the third step in TP, D sends $E_6 = E_{Z(= snx)}(Sig_D, C_D, S_6, T_{D3})$ and $T_{D3}$ to C via a public channel, where $C_D = E_{K5}(m_H, m_B, m_D)$ is encrypted with the help of key $K_5 = h(ID_P\|ID_H\|NID)$. Whereas P decrypts this $C_D$ with the help of $K_4 = h(ID_P\|ID_D\|Y)$ in CP.

But $ID_P$, $ID_D$ is already present in the database of C, and $Y(= sn_x)$ is engaged by C. Therefore, the Privileged insider of C also has all these contents and hence can compute the key $K_4$ in TP.

Consequently, all reports $m_H$, $m_B$, $m_D$ of patients will be revealed by A.

### 4.2. Report Confidentiality:

If these health reports are revealed by the privilege insider of C (i.e. A), then the condition for report confidentiality has been contradicted by these action of A.

(Here the condition for report confidentiality is that the medical reports of P should be accessible by only appointed doctor.)

### 5. CONCLUSION:

When we looked again at Kumar et al's scheme which is with the mutual authentication approach, we concluded that it is not capable of securely transmitting medical reports between patients and doctors. Since these reports are made public by a privileged insider of C, the patient's privacy is violated, and as a result, the reports' confidentially is also ruined. As a result, this scheme doesn't fulfill all the objectives of a secure cloud-assisted Telecare Medical Information System.